\documentclass[twoside]{article}
\usepackage{jwe,epsfig}

\usepackage{graphics, graphicx}
\usepackage[square, comma, sort&compress, numbers]{natbib}
\usepackage{subfigure}
\usepackage{booktabs}
\usepackage{multirow}
\usepackage{paralist}
\usepackage[all]{xy}

\usepackage[hyphens]{url}
\usepackage{hyperref}
\hypersetup{
    colorlinks,
    citecolor=black,
    filecolor=black,
    linkcolor=black,
    urlcolor=black
}

\usepackage[ulem=normalem]{changes}
\usepackage{changes}

\newcommand{\spara}[1]{\smallskip\noindent{\bf #1}}
\newcommand{\mpara}[1]{\medskip\noindent{\bf #1}}

\usepackage{array}
\newcolumntype{L}[1]{>{\raggedright\let\newline\\\arraybackslash\hspace{0pt}}p{#1}}
\newcolumntype{C}[1]{>{\centering\let\newline\\\arraybackslash\hspace{0pt}}p{#1}}
\newcolumntype{R}[1]{>{\raggedleft\let\newline\\\arraybackslash\hspace{0pt}}p{#1}}

\usepackage{verbatim}

\begin{document}
\setlength{\textheight}{8.0truein}

\runninghead{Engineering Crowdsourced Stream Processing Systems}
            {Imran et al.}
      
\normalsize\textlineskip
\thispagestyle{empty}
\setcounter{page}{1}


\alphfootnote
\fpage{1}

\centerline{\bf
ENGINEERING CROWDSOURCED STREAM PROCESSING SYSTEMS}
\vspace*{0.035truein}
\vspace*{0.37truein}
\centerline{\footnotesize
MUHAMMAD~IMRAN$^1$, IOANNA~LYKOURENTZOU$^2$, YANNICK~NAUDET$^2$, CARLOS~CASTILLO$^1$}
\vspace*{0.15truein}
\centerline{\footnotesize\it $^1$Qatar Computing Research Institute,}
\baselineskip=10pt
\centerline{\footnotesize\it Doha,
Qatar}
\baselineskip=10pt
\centerline{\footnotesize\it 
mimran@qf.org.qa, chato@acm.org}
\vspace*{0.1truein}
\centerline{\footnotesize\it $^2$CRP Henri Tudor}
\baselineskip=10pt
\centerline{\footnotesize\it Luxembourg}
\baselineskip=10pt
\centerline{\footnotesize\it ioanna.lykourentzou@tudor.lu, yannick.naudet@tudor.lu}
\vspace*{0.1truein}

\abstracts{
A crowdsourced stream processing system (CSP) is a system that incorporates crowdsourced tasks in the processing of a data stream.
This can be seen as enabling crowdsourcing work to be applied on a sample of large-scale data at high speed, or equivalently, enabling stream processing to employ human intelligence. It also leads to a substantial expansion of the capabilities of data processing systems.
Engineering a CSP system requires the combination of human and machine computation elements. From a general systems theory perspective, this means taking into account inherited as well as emerging properties from both these elements.
In this paper, we position CSP systems within a broader taxonomy,
outline a series of design principles and evaluation metrics,
present an extensible framework for their design,
and describe several design patterns.
We showcase the capabilities of CSP systems by performing a case study that applies our proposed framework to the design and analysis of a real system (AIDR) that classifies social media messages during time-critical crisis events. Results show that compared to a pure stream processing system, AIDR can achieve a higher data classification accuracy, while compared to a pure crowdsourcing solution, the system makes better use of human workers by requiring much less manual work effort.  
}{}{}

\vspace*{10pt}

\vspace*{3pt}

\vspace*{1pt}\textlineskip 

\section{Introduction}
\label{sec:intro}

Stream processing refers to computations performed on an unbounded, high-speed, continuous, and time-varying sequence (stream) of events [49], which typically cannot be stored off-line and needs to be processed on-line. If performed timely and effectively, stream processing can play a vital role in supporting real-time decision making. Many practical applications have been developed based on the stream processing paradigm, including real-time road condition monitoring, event-based business processing and profile-based query processing, implemented through various stream processing frameworks, such as the Aurora, TelegraphCQ, S4 and Storm project system among others ~\cite{neumeyer2010s4,marz_2013_storm,carney2002monitoring,chandrasekaran2003telegraphcq}.

A significant drawback of traditional stream processing systems is that they rely entirely on automated algorithms of data processing, and as such they are limited by the processing capabilities of these algorithms. Indeed, it is often the case that data stream to be processed is imprecise, highly variable and previously unseen, yet the decision based on this data needs to be made in real-time. In this case, automated processing on its own may produce undesirable results, harming the decision-making process it needs to support.
For example, suppose we are interested in monitoring social media to identify messages related to a major crisis event (e.g. an epidemic spreading such as the H1N1 flu pandemic in 2009). The moment this data is issued it needs to be correctly categorized and classified, in order to inform health and policy decisions accurately and on time. Unfortunately, relying only on automated data processing algorithms may not be adequate, firstly because user-generated data in Twitter may partially be erroneous, due to human mistake or intentionally, and secondly because new data may have different characteristics than existing data used to create the algorithms used by the system.
In machine learning, this is referred to as the \emph{domain adaptation problem}, where due to concept drift and other issues, a model trained on one event does not perform well on another. Back to our previous example of epidemic spread detection, a fully automatic  stream processing algorithm trained on data from the H1N1 epidemic, may not function correctly on a new epidemic (e.g. the MERS-coronavirus outbreak in 2014), for instance due to the presence of unique vectors (e.g. spread through camels instead of birds) which have not been seen in previous situations.
Nevertheless, failing to correctly filter the stream data of the new situation may lead to incorrect conclusions.
It therefore becomes apparent that \emph{in cases where critical---in terms of cost, time or other valuable resource---decision-making needs to take place in real-time, based on data streams that are potentially noisy and unseen, fully automated stream processing systems do not suffice}. Similar scenarios to the above include detecting intrusion or fraud based on anomalous or unusual activities.

Crowdsourcing is a recent computing paradigm that has been used to address the above problems~\cite{geiger_2011_crowdsourcing}. It refers to utilizing a large number of users (a crowd) to perform a task, by splitting it into several sub-tasks and allocating them to individual workers. Crowdsourcing presents the significant advantage of involving human intelligence; and with it, the ability of humans to categorize, filter and evaluate, often better than automated methods.
In fact, crowdsourcing has proven to be better than current computer algorithms in numerous cases, ranging from image recognition and audio transcription to real-time document syntax checking. In our example above, crowdsourcing could make use of the ability of humans to make better sense of a textual message than any algorithm, and in this sense help categorize and disambiguate the information included in a crisis data stream much more effectively.
Given that typical crowdsourcing relies entirely on human effort, which inevitably entails the risk for low recognition quality, 
typical crowdsourcing approaches apply redundancy, i.e. they allocate the same task to multiple workers. This redundancy decreases the throughput and increases the cost of crowdsourced processing.
Adding to this problem is the fact that stream processing tasks, unlike other types of input to crowdsourcing, necessitate a very high response rate (throughput) from the crowd. For example, during the Hurricane Sandy a peak rate of 16,000 tweets/minute was observed, and in such case, pure crowdsourcing may easily fail due to human limitations. \emph{Summarizing, while fully automated processing may not be appropriate because of errors or quality issues, fully crowdsourced processing may not be appropriate because of processing time (throughput) and cost.} 

The above problem leads to the nascent class of crowdsourced stream processing systems (CSPs), which combine automatic and human processing elements applied on time-sensitive, often critical information in the form of data streams. Certain instances of such systems have recently appeared, requiring different amounts of automated and human involvement, and involving different architectural and design patterns, of which we provide an overview in the next sections.
Despite their appeal and necessity, however, no concrete framework exists currently to organize the characteristics, functionalities, architectural patterns and problem cases that CSP systems apply on. As a result, the design of CSPs today is ad-hoc and, at best, intuition-based. A systematic formalization and synthesis of these systems could significantly improve the design and engineering of CSP applications, leading to better designs and to performance improvements of current systems, and guiding the development of new ones. Given the added value and criticality of accurate, fast stream processing, and the abundance of applications that need it, one may easily understand why such systematization is essential. To the best of our knowledge, \emph{this is the first work attempting to systematize the description and analysis of CSPs through a concrete functional, architectural and evaluation framework.}

\spara{Contribution and organization of this paper.}
First, we propose a generic framework for the design and analysis of crowdsourced stream processing systems. Our contribution begins in Section~\ref{sec:taxonomy} by positioning CSPs into a broader taxonomy of systems. 
Then we describe design objectives, principles and quantifiable evaluation criteria (Section~\ref{sec:principles}). 
In Section~\ref{sec:framework} we introduce a generic application design framework in terms of composable elements and communication channels. 
In Section~\ref{sec:patterns} we present a series of design patterns to solve specific design problems.
Section~\ref{sec:evaluation-aidr} shows a case study in which we design and analyze a CSP system using our framework. 
Finally, Section~\ref{sec:related-work} outlines related work and Section~\ref{sec:conclusions} summarizes our main conclusions.


\section{Characterizing crowdsourced stream processing systems}\label{sec:taxonomy}

In this section we introduce the main characteristics of crowdsourced stream processing systems (CSPs). We start by providing a systemic view on CSPs, giving their elements, functions and properties. Then, we provide tangible examples of CSPs using a series of systems previously described in the literature.

\subsection{The CSP system}\label{sec:CSP-System}

In order to better understand what a CSP system is, we follow a relaxed systemic-modeling approach~\cite{banathy1997taste}, which borrows explanatory elements from systems science and General Systems Theory~\cite{VonBertalanffyGeneralSystemTheory}. According to this theory, {\em a system is a set of interacting elements} that has {\em emerging properties} which are richer than the sum of the properties of its parts. 

\begin{figure} [h]
\centering
\includegraphics[width=0.4\columnwidth]{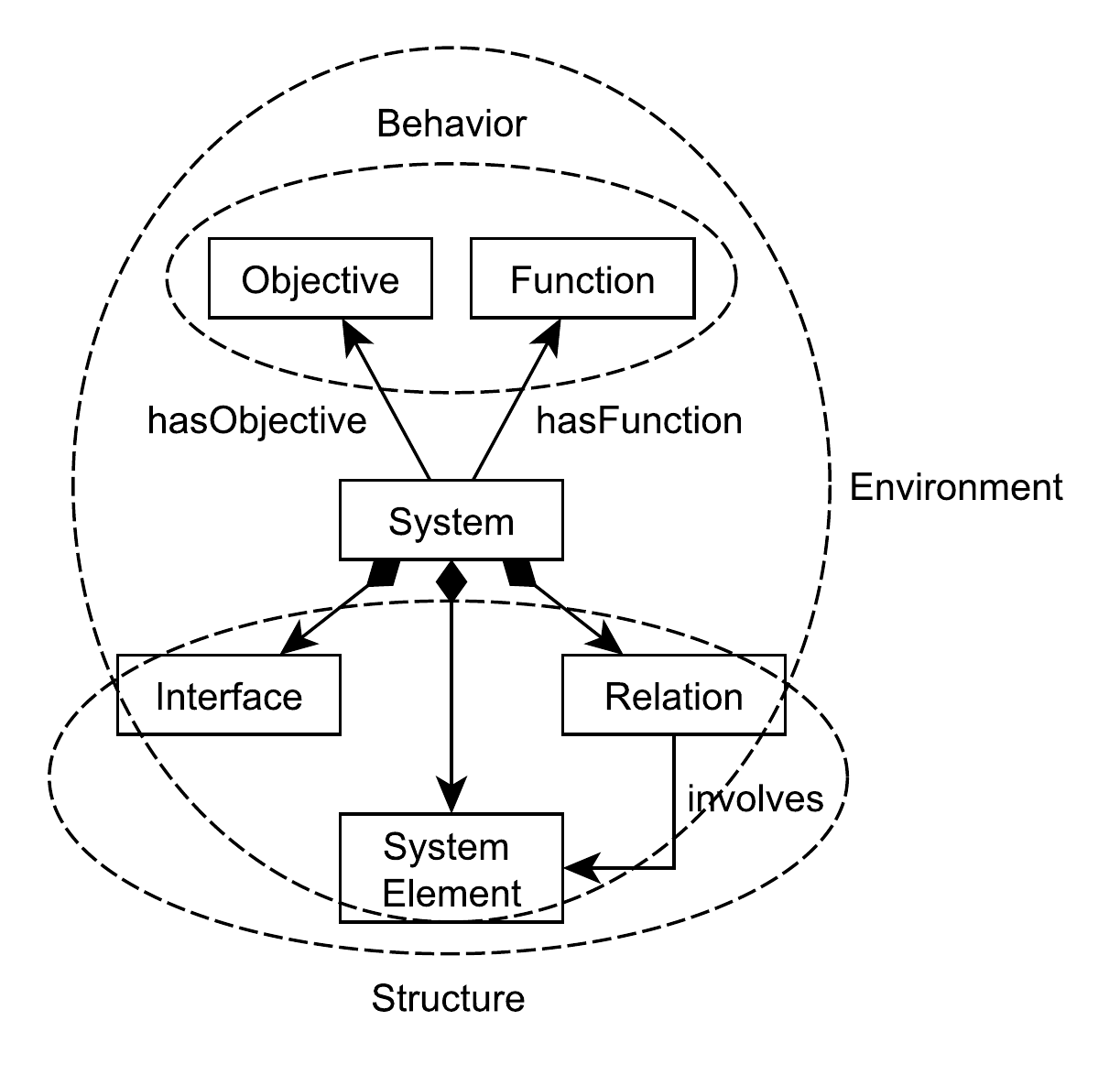}
\fcaption{\label{fig:systemMModel}Simplified meta model of a System. Diamond arrow represents the composition relation.}
\end{figure}

A system can be examined from two main perspectives~\cite{Dietz:2008:EOE:1363686.1363824}: (i) a \emph{behavioral} or \emph{teleological} perspective, where the system's behavior and goals are examined, and (ii) a \emph{structural} perspective, where the system's structure, architecture and operations are considered. These perspectives are supported by notions that include environment, objective, function, element, relation and interface~ \cite{Naudet2010176}, as illustrated by the system model of Figure~\ref{fig:systemMModel}, adapted from~\cite{Naudet2010176}. \emph{Environment} is anything outside the system's boundaries. It influences the system as well as the system influences it. \emph{Objective} is the system's finality at a given time, a notion that directly influences the system's structure and functionality. \emph{Function}  is the set of actions the system can execute in order to realize its objective. \emph{Element} is a component of a system, which can itself be a system. Finally, \emph{Interface} is the element through which the system establishes connections with its environment, each system's element having interfaces through which \emph{Relation}s with other elements are established.
Environment, objective and function provide the \emph{behavioral perspective} of the system, i.e. they denote the way the system acts and reacts. Element and interface provide the \emph{structural perspective} of the system, i.e. they materialize the internal organization and architecture of the system.

A CSP is a combination of two systems: a \emph{Stream Processing system} and a \emph{Crowdsourcing system}. It keeps the objective of the first, while using the other as a tool to better reach this objective.


%

\spara{Stream processing system.} 
From a {\em behavioral perspective}, the objective of a stream processing system is to process data streams.
%
Example applications include data mining \cite{sudhakar2012data}, clustering \cite{aggarwal2003framework,zhou2008tracking}, classification \cite{zhang2012framework}, time series analysis \cite{lin2003symbolic,zhu2002statstream} and burst detection \cite{sun2010burst}, among many others.

From a {\em structural perspective}, a stream processing system is a composition of {\em processing elements} that read an input data stream and write an output data stream. 
Typically, each processing element performs a fairly simple task and is thus easy to implement, debug and maintain. The data passed between elements are data streams as well.
Apart from handling a particular type of data (data streams) and employing single-pass processing, another property of stream processing systems is that---at a structural level---each processing element is assumed to have limited memory/storage, i.e. the amount of data it can store is small in comparison to the total amount of data that passes through it~\cite{muthukrishnan2005data}. 
Stream processing systems have important properties which will be evaluated to measure the system performances: \emph{latency} (the time that it takes for a stream data item to be processed), \emph{throughput} (the number of items processed/unit time), \emph{load adaptability} (the response of the system to abrupt changes in the input load), \emph{processing quality} (processing approximation error and probability of error, in case this system functions on the basis of an approximation algorithm).

\spara{Crowdsourcing system.} 
From a {\em behavioral perspective}, the objective of a crowdsourcing system is to perform some tasks by using the intelligence of a large group of persons, called the crowd. Typically, the goal is to apply human intelligence to  tasks that cannot be performed effectively by fully automated means. This is done by involving large numbers of workers who perform small portions of a larger task, improving the efficiency of more traditional ways of work organization by reducing completion time and cost. 


From a {\em structural perspective}, a crowdsourcing system is composed of two subsystems interacting one with the other: an information system and the crowd. While the crowd brings along the power of its numbers and a large diversity of knowledge backgrounds, it still remains a human system that can have unpredictable behavior. Thus the crowd's influence on the crowdsourcing system is usually limited to the expected input of the former to the latter, an input that is moderated or filtered when necessary. When the crowd is given more freedom that can influence the functionality of the whole crowd system, more complex regulation mechanisms are put in place~\cite{naudet2014_smap_submitted}.


At a more detailed level, crowdsourcing systems vary significantly in their structure (for comprehensive surveys see ~\cite{yuen_2011_crowdsourcing,doan2011crowdsourcing}). These variations depend on the application (knowledge or creativity-intensive or more mechanistic), type of input data (homogeneous or heterogeneous) and coordination mode (independent workers, collaboration, competition, mixed). Also, the type of task handled varies greatly from one crowdsourcing system to the other. Among the tasks that workers can execute, we may find:  \emph{providing input}, \emph{correcting}, \emph{validating} a given information, \emph{comparing}, \emph{categorizing}, but also \emph{searching} for information, \emph{synthesizing} knowledge and \emph{providing judgments}, among others. This leads to prototypical classes of crowdsourcing tasks: binary classification, n-ary classification, and data entry contributions. 
In \emph{binary classification}, workers are asked to give a ``yes/no'' answer to a given question, such as ``do these suspicious activities constitute an attack?'' \cite{debar2001aggregation}, ``does this entity correspond to this word in this context?''~\cite{demartini_2012_zencrowd}, ``do these two records correspond to the same person?''~\cite{whang2013question}.
Examples of \emph{n-ary classification}, include
determining if a word in a sentence represents a person, a place, or an organization~\cite{finin2010annotating},
determining if an emergency-related tweet describes infrastructure damage, casualties, injuries, needs, etc.~\cite{imran2013extracting}, 
determining which emotion corresponds to a tweet~\cite{liu2012cdas}, performing relevance assessments ~\cite{alonso2008crowdsourcing}, determining certain properties of a photo of a galaxy~\cite{kamar_2012_galaxyzoo}, etc.
In \emph{data-entry contributions}, workers are asked to perform a data-entry task such as:
``rewrite this sentence in a shorter way''~\cite{bernstein2010soylent},
``fill in the missing fields in this record''~\cite{franklin2011crowddb},
``make a drawing of an articulated figure''~\cite{demartini_2012_zencrowd}, or
provide free-text labels for an image~\cite{von2004labeling}.
Binary and n-ary classification tasks are usually more restricted, self-contained and can be more easily split to micro-task level, compared to data-entry ones. 
 
Many real-time crowdsourcing applications (e.g. \cite{franklin2011crowddb, geiger_2011_crowdsourcing}), are designed to have low latency, and to rely on homogeneous input data, highly decomposable and self-contained tasks, parallelizable task handling and centralized work coordination. This kind of crowdsourcing is referred to as {\em crowd processing}~\cite{geiger_2011_crowdsourcing}, and is the one that can be most easily integrated with stream processing systems to create CSPs.



As crowdsourcing is part of a CSP, key challenges of it are inherited by CSPs including dealing with the innate uncertainty of human work (e.g. workers might leave unexpectedly and their skills might vary~\cite{roy2013crowds}), and the fact that crowdsourcing is in general slower and more costly per item than fully automatic processing. Then, the crowdsourcing properties that will be usually of concern in the design of a CSP are the crowdsourcing \emph{cost} (the cost of the work of the people providing the human intelligence) and the \emph{quality} of the crowd input.



\spara{Crowdsourced stream processing system (CSP).}
CSPs lie in the junction between automated stream processing systems and crowd processing systems. They inherit behaviors and structural elements from both, while demonstrating its own emerging properties.
What CSPs inherit from stream processing systems is their general objective: to process input streams efficiently. What differentiates them is the \emph{class of problems} that CSPs can handle.
From crowdsourcing systems, CSPs inherit the capability of executing data processing tasks that require human intelligence, in contrast to the tasks traditional stream processing systems execute, which can be done by computer algorithms alone. In addition, CSP inherits the uncertainty/variability introduced by human work, and its costs, which necessitates methods and design choices to mitigate it.

From a \emph{structural perspective} a CSP is composed of processing elements (PE), which involve either human or machine elements: crowdsourced PE (CSPE) and automatic PE (APE). These elements can be combined in different ways. We may distinguish two prototypical connection topologies, which generate three possible kind of relations:
\begin{compactitem}
	\item \emph{Parallel.} In parallel connections, both automatic and crowdsourced processing elements function on the same type of task and layer of data stream processing, therefore such tasks have no dependencies. 
	\item \emph{Serial.} In serial connections, tasks have dependencies among them, requiring a certain degree of serialization in their processing. 
	\item \emph{Hybrid.} We remark that in general for solving real-world problems, hybrid topologies are used, involving a combination of parallel and serial connections.
\end{compactitem}

In general, in a CSP the combination of automatic and crowdsourced processing elements is such that it \emph{avoids having crowdsourced processing elements in high-volume critical paths}, employing them only in non-critical paths or in low-volume areas of a critical path. The volume of data to be processed by crowdsourced processing elements is often reduced by automatic processing elements performing filtering or aggregation operations.
  
From a \emph{behavioral perspective} the objective of CSPs is to process data streams. 
This is supported by different functions given to the APE or CSPE elements, which we detail in the following.
Some tasks that workers can execute within a CSP include \emph{providing input} to other processing elements, \emph{correcting} an element's output, \emph{validating} that an  element performed as expected, and \emph{training} the automated processing elements so that they can learn to perform a task.
%
The most suitable tasks for crowdsourced processing elements are those that involve \emph{crowd processing} (such as binary and n-ary classification), while open-ended tasks are harder to integrate in a CSP because they may not be self-contained and are often non-decomposable. 

\begin{table}[t]
\centering
\tcaption{Types of processing elements CSP systems contain, with relations and some indicative functions.}
\label{tab:systemicCSP}
\small\hskip-0.5em\begin{tabular}{p{2.5in}L{1.4in}} \toprule
Type of Processing Elements & Functions  \\\midrule
\multirow{5}{*}{Automatic Processing Element (APE)} & Computation \\
	& Filtering \\
	& Task generation \\
	& Task assignment \\
	& Task aggregation \\ \midrule
\multirow{3}{*}{Crowdsourced Processing Element (CSPE)} & Binary classification \\
	& N-ary classification \\
	& Data entry \\
\bottomrule
\end{tabular}
\end{table}

Whereas crowdsourced processing elements deal with higher-difficulty tasks, automatic processing elements are employed for efficiency. Indicatively, we can distinguish between five functions for the APE (see Table~\ref{tab:systemicCSP}). First, and perhaps most frequently, automatic elements can be used to perform arbitrary \emph{computations} over the data (transform, aggregate, sort, join, etc.~\cite{mendes_2009_performance}). 
%
Secondly, automatic elements can be used for task \emph{filtering}: performing sampling to reduce the number of elements that need to be processed. This may help to reduce the number of crowdsourcing calls by selectively using crowdsourcing only for difficult items (e.g. \cite{kamar_2012_galaxyzoo,demartini_2012_zencrowd}).
%
Third, automatic elements can be used for automatic \emph{task generation}, as in the case of systems that need to convert input streams into questions for humans (e.g. graph searching in~\cite{parameswaran2011human}, crowd-assisted data joins in~\cite{franklin2011crowddb}).
Fourth, automatic elements can be used for \emph{task assignment} purposes to ensure each crowdsourcing worker receives a task s/he is most capable of doing, e.g., with the online algorithm described in~\cite{ho_2012_task_assignment}.
Finally, automatic elements can be used for \emph{aggregation} of the output of other processing elements, for instance in order to ensure output quality through redundancy, as in the ``get-another-label'' system~\cite{ipeirotis_2010_get_another_label}.

In section~\ref{sec:framework}, we will provide a prototypical structure of a CSP and detail more precisely APE and CSPE and the way they function together.





\subsection{Examples}\label{sec:examples}

We close this section by illustrating three indicative CSPs examples, categorized according to the functions of the automatic processing and crowdsourced processing elements and the kind of relation between them (Table~\ref{tab:taxonomy_examples}).

\begin{table}[t]
\tcaption{Three indicative examples of CSPs.}
\label{tab:taxonomy_examples}
\small\hskip-0.5em\begin{tabular}{p{1.3in}L{1.4in}L{1.6in}L{0.7in}} \toprule
Example CSPs & Crowdsourced processing & Automatic processing & Dominant composition\\\midrule
\mbox{Network intrusion} \mbox{detection}~\cite{om2012hybrid}
& input; binary classification
& filter
& serial \\
\mbox{Automatic online} \mbox{classification}~\cite{whang2013question}
& training; n-ary classification
& task generation
& hybrid \\
\mbox{On-demand missing} \mbox{data completion}~\cite{franklin2011crowddb}
& input; comparison
& task generation, aggregation
& hybrid \\
\bottomrule
\end{tabular}
\end{table}


\spara{Network intrusion detection.} Automatic processing elements monitor network or system logs, to detect candidate malicious activities and generate alarm reports about possible attacks~\cite{om2012hybrid}. Human input is then needed to manually verify candidate alarms and take further action (perform remedy, re-tune the system) in response. Processing composition in this example is serial, with the automated element performing the filtering part (detect) and the human element performing binary classification (verified attack or not). 

\spara{Automatic online classification.} Crowdsourced processing elements are asked to correct n-ary classification of entities (e.g. mentions of people or places in a text) into semantic clusters, helping at the same time to train the automated element of the CSP into performing more accurate automatic classifications~\cite{whang2013question}. In this case, datasets can be static or dynamic (query-dependent). The processing composition in this system is hybrid (namely, a loop), with the automated element generating entity resolution questions, humans answering them and then these answers being sent back to the automated elements for training and decision purposes. 



\spara{On-demand missing data completion.} Crowdsourced processing elements are orchestrated to fill in missing data with respect to a query~\cite{franklin2011crowddb}. Among other operations, in this system crowdsourcing workers complete missing data in tuples that need to be selected or compared with other tuples in response to a query. Queries are not known beforehand, and the underlying dataset can change over time.


\begin{table*}
\tcaption{Design objectives and principles}\label{tab:principles}
\small\begin{tabular}{p{1.0in}L{1.4in}L{1.4in}L{1.4in}}\toprule
                 &        & \multicolumn{2}{c}{Design principles}\\
Design objective & Example metric & Automatic  components & Crowdsourced components \\ \midrule
Low Latency      & End-to-end time            & Keep-items-moving              & Simple tasks \\
High Throughput  & Output items per unit of time & High-performance processing & Task automation \\ 
Load Adaptability& Rate response function     & Load shedding, load queueing  & Task prioritization   \\
Cost Effectiveness& Cost vs. quality/throughput/etc. & --- & Task frugality \\
High Quality     & Application-dependent & \multicolumn{2}{l}{~~~~~~Redundancy/aggregation and quality control} \\
\bottomrule
\end{tabular}
\end{table*}

\section{Design objectives, principles and metrics}\label{sec:principles}

In the previous section we presented a characterization of crowdsourced stream processing systems. In this section we give design principles that help building efficient systems. This efficiency can be achieved through optimising the important properties of CSPs inherited from its parents: stream processing and crowdsourced systems (see section~\ref{sec:CSP-System}). Each resulting design objective is quantifiable through a certain metric, and trade-offs among different objectives may occur, as summarized on Table~\ref{tab:principles}.

The first three objectives (low latency, high throughput and load adaptability) are common in the stream processing literature~(e.g. \cite{turaga2010design,mendes_2009_performance}); while they are not crowdsourcing-specific, we adapt them to the crowdsourced stream processing setting. We do so by indicating how to address each objective in the automatic and crowdsourced components of the system.
The last two objectives (cost effectiveness and high quality) are a consequence of the presence of humans as part of the system. 

\subsection{Low Latency} \label{subsec:principles-latency}

Latency is the time it takes for an item to be processed, and a low latency system takes less time to deliver results than a high latency system.
In the literature, low-latency crowdsourcing stream processing systems have been referred to as real-time crowdsourcing systems (e.g. \cite{kittur2013future,bernstein_2011_twoseconds}).

\spara{Measuring latency.}
A typical measure for the latency of a system is the average latency of the items traversing through it.
The latency of an item is trivial to compute if the mapping from input to output items is one-to-one. If that is not the case (e.g. because each output item depends on several input items that arrived at different times), then for each input item that contributed to one or more output items, its latency can be measured as the time it took for the first such output item to be produced. 

\subsubsection{Automatic components: keep data moving}

Automatic components should avoid unnecessary latencies due to network delays, storage operations, or other causes. Input items may be out-of-order, missing or delayed. Automatic processing elements should use non-blocking operations, never waiting indefinitely for some data items to arrive before continuing processing~\cite{stonebraker20058}.

\subsubsection{Crowdsourced components: simple tasks} \label{subsec:trivial-tasks}

Task design depends on many aspects that include incentives, interface, task description, and more importantly the task itself~\cite{kittur2013future,finnertykeep}. Research and practice suggest to decompose complex tasks into simple subtasks, where each task is designed to be simple and follow specific requirements. This can not be understated, as it helps both to reduce the cost and to increase the quality of work.
Simpler tasks are completed earlier~\cite{horton_2010_economics}, not only because each task is completed faster, but because the pool of people with the skills required to complete a task is larger if the task is simpler. A difficult task may introduce latency, and it may also reduce output quality~\cite{kazai_2010_influence}.

Task decomposition helps reduce individual task complexity.
%
The latency of the decomposed task may be lower if each individual task can be answered faster and/or by a potentially larger group of crowdsourcing workers.
We note that task decomposition may have different constraints across applications. While in general tasks should be as simple as possible, in some cases it may be efficient to bundle a few sub-tasks together, e.g. to reduce context switching and reduce latency.

\subsection{High Throughput}

Throughput is the speed at which items are processed: a high-throughput system can process more items per unit of time than a low-throughput one. The throughput of a system is a function of its design and of the throughput of its components. 

\spara{Measuring throughput.}
Throughput can be measured as output items per unit of time, e.g. items/second. 

\subsubsection{Automatic components: high performance}

A stream processing system should process and accommodate long-running analysis requirements almost in real-time~\cite{turaga2010design}. Each automatic processing element must be implemented to have high performance, and the application infrastructure including communication channels must be able to have a high throughput.

\subsubsection{Crowdsourced components: task automation}

In a crowdsourced stream processing system, in general the throughput of the crowdsourced processing components is lower than that of the crowdsourced processing components. Crowdsourced components may become a bottleneck.

Automating as much work as possible is a way of maintaining a high throughput: any aspect that can be automated should be not passed to a crowd. For instance in binary classification if we are searching for items belonging to a positive class, if an item can be automatically classified into the negative class, it does not need to be given to crowdsourcing workers.

\subsection{Load Adaptability}

The rate of input data may experience sudden changes including significant bursts. 
Adaptability is the capacity of the system to respond to a surge in demand.

\spara{Measuring load adaptability.}
Adaptability can be measured as the response function of throughput and latency vs input load. Ideally, these variables should not be strongly affected by increases in input load. In other words, we should not observe a significant reduction in throughput, or a significant increase in latency.

\subsubsection{Automatic components: load shedding/queuing}

Surges in input rate may require to use {\em load shedding}, this is, completely ignoring data items that are beyond the processing capabilities of a processing element at a given time~\cite{tatbul2003load}. Buffering may prevent shedding by allowing {\em load queuing}, using a bounded-size queue.

\subsubsection{Crowdsourced components: task prioritization}

The system should prevent crowdsourcing from becoming a bottleneck. This can be done by having a method for prioritizing tasks, in such a way that during periods of increased input rate, more tasks are taken by automatic parts of the system than normally. This may be done at the expense of output quality, if necessary.

For instance, \citet{demartini_2012_zencrowd} pass to the crowd only tasks (in their case, entity linking tasks) for which an automatic system is uncertain, i.e. it has not given with high-confidence a positive or a negative answer. The thresholds of what constitute a high-confidence answer could be tunned to be able to handle more input items per unit of time.

\subsection{Cost effectiveness}

Given that crowdsourcing work is neither unlimited in supply nor free (even when done by volunteers, their time and motivation are precious resources), a principle of {\em task frugality} needs to be applied.

Crowdsourcing is usually compensated in proportion to the time spent by workers. Cost is therefore a function of (i) the payment per unit of time, (ii) of the number of items to process, (iii) the time needed by each worker to complete one item, and (iv) the plurality of workers per item. Effort is also a parameter, eventually translatable to cost as well.
Increasing the payment per unit of time seems to have little effect in work quality, but can help reduce its latency~\cite{mason_2010_financial}.

The number of items to process is related to task automation---replacing crowd processing by automatic processing when possible---but also to the way processing elements are composed.
Similarly to query planning in traditional database systems, heuristics or optimization methods can be used to determine the order in which automatic and crowdsourcing operations need to be done~\cite{franklin2011crowddb}. Task workflow design can be optimized through automatic processing elements, as described by~\citet{dai_2013_pomdp}.

The time and effort needed by each worker to complete one item can be decreased by decreasing task subjectivity and difficulty, for instance by decomposing into smaller sub-tasks, which may also lead to lower latency (Section~\ref{subsec:trivial-tasks}). Task completion time can also be reduced by using more efficient worker-to-task allocation schemas, e.g., based on the average completion time and skills of each worker~\cite{boutsis_2013_crowdsourcing}. 

The plurality of workers per item reduces the effect of cheaters/spammers by using aggregated labels. This can be optimized by using a cost-sensitive objective, e.g., combining cost with quality as in~\cite{gao_2013_online}.

\spara{Measuring cost effectiveness.}
This can be done in comparison with other aspects, e.g. cost/latency, cost/quality, etc. For instance, in~\cite{fan2013hybrid} it is shown how larger budgets (in their case, fraction of items that are crowdsourced) yield better overall accuracy. The time to complete a task is also reduced when payment is increased~\cite{mason_2010_financial,heer_2010_graphical}.

\subsection{High quality}\label{subsec:principles-quality}

The addition of human elements makes crowdsourcing stream processing non-deterministic. Crowds vary in their composition and individual workers may provide varying levels of quality.

Output quality can be maintained by the interaction between automatic and crowdsourced components through operations such as aggregation, quality management and cheating detection (as mentioned e.g., in~\cite{yuen_2011_crowdsourcing} and others). Section~\ref{subsec:qa-loop} describes a design pattern (``quality assurance loop'') based on this type of interaction. Furthermore, the automated elements can play an important role in regards to incentives engineering, for example in diversifying task recommendations (bored workers perform worse than interested ones~\cite{kazai:an}), as well as in allocating incentives to tasks \cite{Yang:2013:IT:2510649.2511161}, with an aim to attract qualitative worker contributions and thus increase overall task quality.

\spara{Measuring quality.}
Quality should be measured in an application-dependent manner. The metric may be binary (correct item vs. incorrect item) or continuous, and it may not be a single scalar but comprise several aspects.


\section{General CSP Framework} \label{sec:framework}


Taking into account the characteristics of CSP systems and the design principles presented above, in this section and the following one, we describe a framework for their design.
%
In accordance with the CSP system model provided in section~\ref{sec:taxonomy}, this framework allows specifying the elements of a CSP system and communication flows among them, in a standardized way that allows to easily create various kinds of topologies, and to modify them when required. In the following we detail the proposed CSP design framework from a {\em system-level} (Section~\ref{subsec:framework-applications}), {\em processing elements} (Section~\ref{subsec:framework-pe}) and {\em communication flows} perspective. (Section~\ref{subsec:framework-com}).

\subsection{System-level technical CSP meta-model}\label{subsec:framework-applications}

\begin{figure}[t]
 \centering
    \includegraphics[width=0.95\columnwidth]{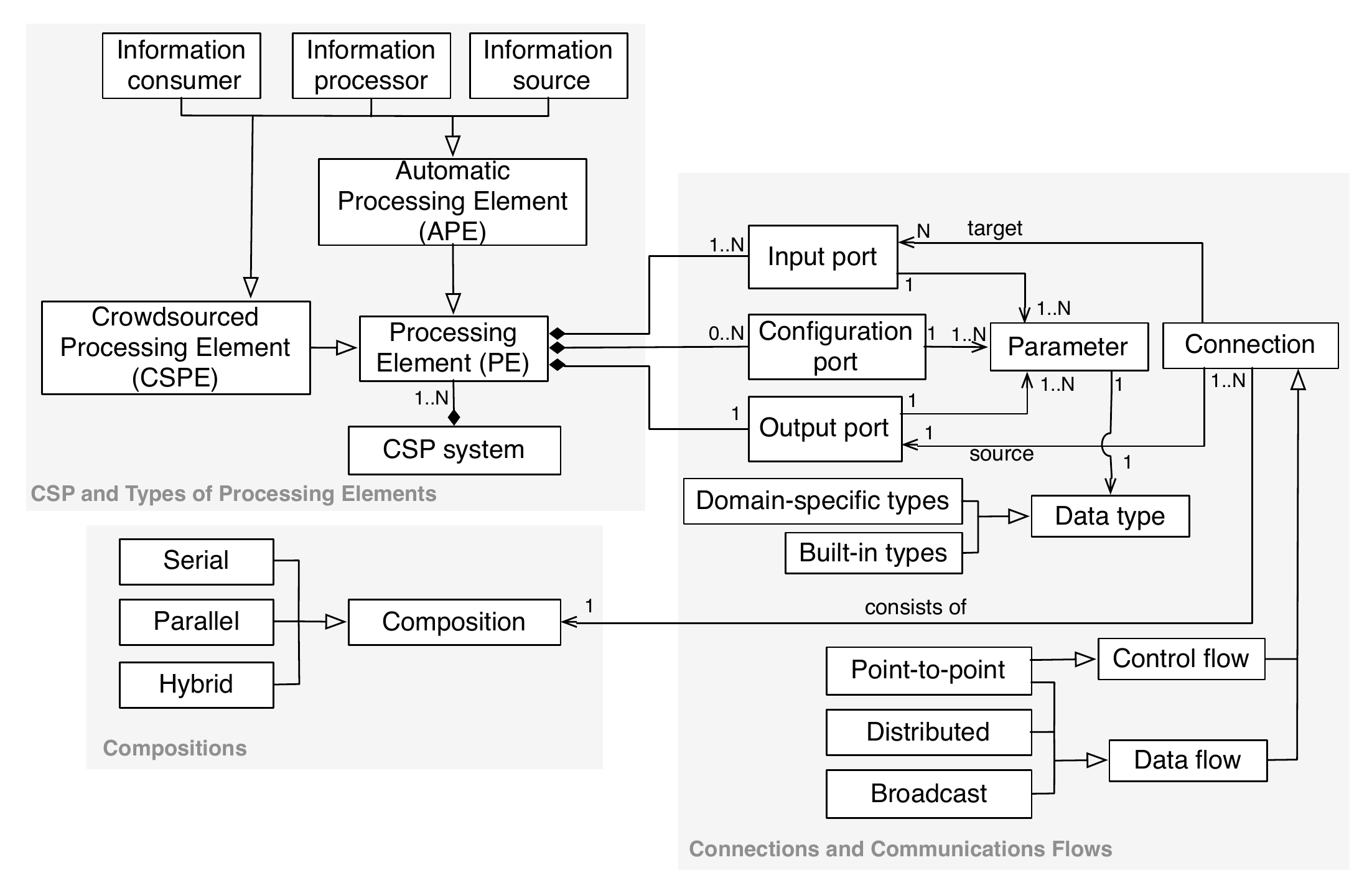}
  \fcaption{Meta-model of a CSP system. The diagram depicts the model of processing elements, connections and communication flows between PEs and composition types.}
  \label{fig:dsp-mm}
\end{figure} 

While the systemic model provided in section~\ref{sec:taxonomy} offers a high-level view, Figure~\ref{fig:dsp-mm} of this section provides a technical meta-model for CSP systems. This figure follows the classical application-level description of software systems, with focus on stream processing systems (e.g.~\cite{neumeyer2010s4,gedik2008spade}). According to this, the basic building blocks of stream processing systems are well-defined software components, which perform various kinds of jobs, and can be connected/composed together.
Similarly, in the CSP meta-model we find the core elements of a CSP system: one or more processing elements~\emph{(PE)}, which can be computer-driven or human-driven (resp. \emph{APE} or \emph{CSPE}).
Each PE performs dedicated tasks, and it may depend on other processing elements in terms of data requirements. The PEs are connected through communication flows for data or control, which are modeled using the \emph{Data flow} and \emph{Control flow} concepts.
The data flow connections tie {\em input ports} and {\em output ports}. Data items are emitted from their source through a (usually single) output port and they are ingested by the target through (possibly multiple) input ports.  
Section~\ref{subsec:framework-com} explains this asymmetry. 
Control connections can also be established to coordinate the behavior between pairs of PEs using (possibly multiple) configuration ports.


\subsection{Processing elements}\label{subsec:framework-pe}

A processing element consumes data items and control flow signals through input ports, implements an application-specific requirement, and emits the processed data items through an output port. There are two types of processing elements as explained below.


\subsubsection{Automatic Processing Elements}

An automatic processing element (APE) is a standard component of stream processing systems, e.g. in~ \cite{neumeyer2010s4,gedik2008spade}. It executes a set of operations in a fully-automated manner on its input stream. Following the streaming computation model~\cite{muthukrishnan2005data}, we assume APEs do not have enough memory to hold all the items that go through them.

\subsubsection{Crowdsourced Processing Elements} \label{sec:managed-crowds}

A crowdsourced processing element (CSPE) employs a large group of people (i.e. a crowd) to process data.
The processing of one item in CSPEs is assumed to be more expensive and slow than in APEs.
There are two ways in which this cost can be incorporated in the design of a system. A first option is to assume that the system can process at most $k$ data items through CSPEs, i.e. there is a fixed \emph{budget} for the crowdsourcing tasks that the system can do. A second option is to consider that there is a certain \emph{cost per task} to be paid, and the system should be designed to minimize that cost.
A CSPE may be implemented through the API of a local crowdsourcing platform (e.g. PyBossa)\footnote{\url{http://dev.pybossa.com/}} or through the API of a remote crowdsourcing application (e.g. Amazon's Mechanical Turk\footnote{\url{https://www.mturk.com/}}, CrowdFlower\footnote{\url{https://www.crowdflower.com/}}, or others).

A processing element (PE) can act as information source, processor, or consumer as explained below.

\spara{Information source.} Processing elements with no input ports are called {\em information source} elements, and perform {\em edge adaptation}~\cite{turaga2010design}, i.e. they convert external stimuli into data to be consumed by the CSP system. Information source PEs are responsible for emitting data continuously. An example of an automatic PE (APE) as information source is a sensor (e.g. seismometer that detects earthquakes) which gathers and provides continuous access to data. An example of a crowdsourced PE (CSPE) as an information source is the data generated through web-clicks of humans on the web.

\spara{Information processor.} Processing elements with both input and output ports are called {\em information processors}.\footnote{In the Storm system~\cite{marz_2013_storm} input sources are named ``spouts'', and information processors and consumers, ``bolts''.}
An example of an APE as information processor, is an example of a computer program that performs some computation over data (e.g. a sort program). An example of a CSPE as information processor is the case of human-processed information (e.g. classification of a twitter message into positive or negative sentiments).

\spara{Information consumer.} Processing elements with no output ports are called {\em information consumers}. Processing elements that only consume information---for example to preserve or visualize it---are known as information consumer elements. An example of an APE as information consumer, is a visualization component that ingests data only to visualize it. An example of a CSPE of type information consumer is a decision-maker (a human) that ingest data from computer program to make a decision during a crisis situation.

\subsection{Connections and Communication flows}\label{subsec:framework-com}

In this section, first, we describe the general communication behavior using channels and ports, second the communication modalities (i.e. {\em point-to-point}, {\em distributed} and {\em broadcasted}) and third the flows among PEs (i.e. {\em control flow} and {\em data flow}).

%

\subsubsection{Channels and ports}

Communication between processing elements is done through generic channels (sometimes referred to as {\em streams}). The concrete implementations of these channels may vary. In case buffering is required, the communication channels may be implemented using bounded-memory queues. If buffering is not required, other message-passing patterns can be used.

The processing elements have input, configuration, and output ports. The input ports are used to ingest data from different data sources. The configuration ports provide a way to configure a PE or to set default values for some of its parameters. The output port is used to emit the output data of the PE. Each port can have multiple parameters and the data types of those parameters can be set using either built-in types (primitive data types) or domain-specific types.
%

%
%

\subsubsection{Communication modalities}

A single output port is all that is needed for most applications to emit output data. However, in addition to the one-to-one communication between two PEs, it is possible that output data has to be distributed from one source to many target PEs in parallel. To this end, the complexity of determining the destination PEs for each output item can be controlled by one of the following {\em communication modalities}:

\spara{Point-to-point.}
Point-to-point communications are the simplest case and have a single processing element as producer and a single processing element as consumer. Essentially, both source and target components in a point-to-point communication modality comprise one port each (i.e. source with the output port and target with the input port).

\spara{Distributed.}
The distribution communication modality supports cases where the output data of a processing element (e.g. data source PE) has to be ingested by multiple PEs. In such cases, multiple processing elements are subscribed to the same channel using the {\em distributed} communication model, possibly filtering data according to certain criteria or keys (as for example in ~\cite{neumeyer2010s4}). The data items are distributed to multiple consumers.

\spara{Broadcasted.}
This communication modality is used when multiple processing elements need to process the same data items at the same time. The {\em broadcasting} communication model duplicates data items to all consumers that are subscribed to a channel.


\subsubsection{Data and control flows}

\spara{Data flows} pass data items across processing elements. 
Data flows should use high-bandwidth channels with some amount of buffering, to provide better load adaptability.

\spara{Control flows} allow a processing element to query or control the behavior of another processing element~\cite{bockermann2012processing}. This can be used, for instance, to modify a parameter or a set of parameters in the operation of a processing element.
Control flows should use low-latency channels with little or no buffering for faster response. Additionally, control flows should typically be used in point-to-point modality since the control signals may depend on the target processing element being addressed.


\subsection{Composition types}
A composition represents a specific topology used to connect various PEs in order to solve a specific problem. CSP systems can be composed using mainly three composition types, as follows.

\spara{Serial composition}. A serial composition represents two or more PEs connected in a serial way, according to which the target PE always depends on the source PE in terms of data or control signal(s).

\spara{Parallel composition}. Parallel composition represents the case where the connected PEs work in parallel and are independent of each other in terms of data or control signals. 
 
\spara{Hybrid composition}. Finally, the hybrid composition represents a mix of serial and parallel types and often a more complex composition of PEs.

\section{Design Patterns}\label{sec:patterns}

The preceding section describes the generic technical components to implement when designing a CSP system. In this section we introduce specific design patterns for the connection of automated and crowdsourced processing elements.
The idea of design patterns 
has been pivotal to software engineering since the work of \citet{gamma_1994_design}.
The design patterns presented in the section provide solutions to common problems with the CSPs context. \emph{By its nature a list of design patterns is always open}, as new problems and new solutions to existing problems can be discovered and incorporated.

In the remainder of this paper, we make use of the following notation: rectangles represent automatic processing elements, rounded rectangles are crowdsourced processing elements, solid lines are data flows, and dashed lines are control flows.

\subsection{Quality assurance loop}\label{subsec:qa-loop}

\noindent{\bf Schematic structure:}
\begin{figure}[ht]
\centering\vskip-1em
\includegraphics[width=0.40\columnwidth]{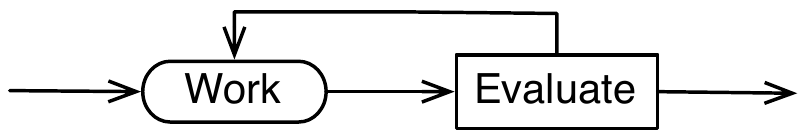}
\vskip-1.5em
\end{figure}


\spara{Problem.}
Given that the quality of work varies across workers, crowdsourcing is almost invariably used with some degree of redundancy. A fixed level of redundancy (e.g. ``each task has to be done by 3 crowdsourced workers'') may not be an optimal solution, given that some tasks may induce larger inter-worker agreement, thus requiring less redundancy, while other tasks may create larger disagreement and require more redundancy.

We would thus like to have per-task guarantees. In some cases, we would like to adaptively ask more workers to perform a task, until a certain criterion is satisfied. In other cases, we may want to ask for more expensive (but more accurate) workers to perform a certain task when we detect that regular workers are not agreeing enough. Some current crowdsourcing platforms offer this feature, by allowing to define the accuracy or expertise level of the involved workers.

\spara{Solution.}
Use an automatic quality assurance loop to aggregate and evaluate crowdsourcing work. The evaluation component may compute cost/quality trade-offs and stop when marginal accuracy increases are smaller than the cost of getting an extra label (as in~\cite{gao_2013_online}). The evaluation component may also keep a model of the trustworthiness of each individual worker, based on his/her past level of agreement (as in ~\cite{ipeirotis_2010_get_another_label}).

\spara{Applicability.}
This pattern is applicable when the quality of the crowdsourcing work can be evaluated automatically by cross-checking labels from different workers. This pattern also applies when there is a global constraint that the output should honor, for example, if we are trying to obtain a total order on items that are being compared pairwise by workers. In this scenario, the evaluation component has to detect violations on the output (a violation of the transitivity property in this example), and ask for more crowdsourcing work to be done on the problematic subsets of the data.

\subsection{User-specific task assignment}\label{subsec:task-assignment}

\noindent{\bf Schematic structure:} 
\begin{figure}[ht]
\centering\vskip-1em
    \includegraphics[width=0.40\columnwidth]{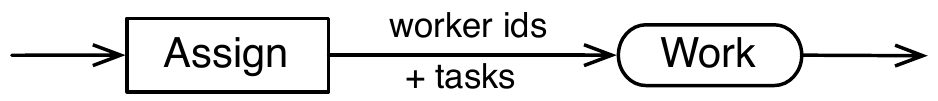}
\vskip-1.5em
\end{figure}


\spara{Problem.} The crowd members are heterogeneous in their capacity to perform tasks. Assigning the ``right'' task to a worker leads to a reliable output, but failing to do so leads to an unreliable one.

\spara{Solution.}
An automatic processing element maintains a model of the skill of workers for different tasks. When feeding a task to a crowd, it indicates what is the specific worker-id that must perform each task.
This design pattern should be combined with a quality assurance loop.

\spara{Applicability.}
This pattern is applicable when there is a variety of different skills needed for a given crowdsourcing job (comprising multiple tasks) and it is possible to automatically compute an estimation of the quality of the per-skill output of a worker performing a certain task (see for example \cite{Goel:2014:ATW:2567948.2577311} where workers of different sets of skills are matched to heterogeneous tasks through a mechanism design-based automatic element). 


\subsection{Process automatically, verify manually}\label{subsec:process-verify}

\noindent{\bf Schematic structure:} 
\begin{figure}[ht]
 \centering\vskip-1em
    \includegraphics[width=0.40\columnwidth]{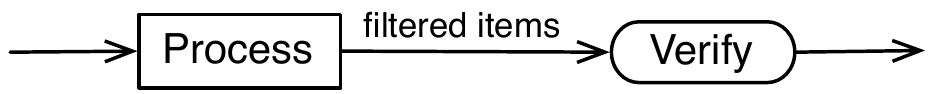}
\vskip-1.5em
\end{figure}

\spara{Problem.}
The input throughput exceeds the budget of crowdsourcing calls, so the data needs to be reduced before being crowdsourced.

\spara{Solution.}
An automatic processing element operating as a ``detector'' can act as a filter, and passes only the elements that are over a certain criterion towards a ``verifier'' crowdsourced processing element.
For instance, crowdsourced content moderation (for profanity or hate speech or inappropriate content as in \cite{myhuemcgowran2012}) could pass only suspicious messages, containing certain keywords or image features, to a crowd of workers. 

\spara{Applicability.}
In the context of binary classification tasks (other cases can be dealt with in a similar manner), this pattern is useful when there are methods to filter-out true negatives deterministically or with high precision. In this case, the number of crowdsourcing calls can be reduced by passing to the crowd only the examples that have a sufficient probability of being positive.

\subsection{Supervised learning}\label{subsec:supervised-learning}

\noindent{\bf Schematic structure:} 

\begin{figure}[ht]
 \centering\vskip-1em
    \includegraphics[width=0.40\columnwidth]{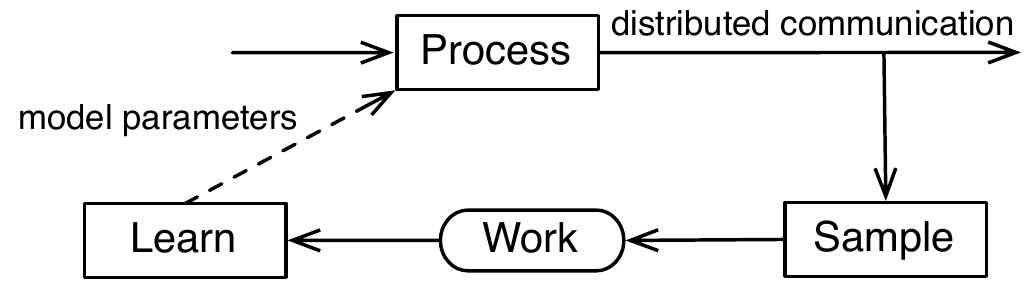}
\vskip-1.5em
\end{figure}

\spara{Problem.}
The input throughput vastly exceeds the budget of crowdsourcing calls, so an automatic system needs to learn to mimic the work process performed by crowdsourced workers.

\spara{Solution.}
An automatic processing element runs a parametrized machine-learned model. Its output is sampled according to a certain criterion, and sent to a crowdsourced processing element to provide {\em training labels}. These labels are used to learn a new model, which is sent through a control signal to the main automatic processing element.
The sampling can be done uniformly at random, or following the idea of {\em active learning} to maximize the gains in accuracy for every extra label.
An example of this solution is \citet{kamar_2012_galaxyzoo}, who use machine vision to identify galaxies based on models learned from human labels.

\spara{Applicability.} 
This pattern can be applied when it is possible to learn an automatic model of the process the crowd performs.

\subsection{Crowdwork sub-task chaining}\label{subsec:subtask-chaining}

\noindent{\bf Schematic structure:} 
\begin{figure}[h!]
 \centering\vskip-1em
    \includegraphics[width=0.70\columnwidth]{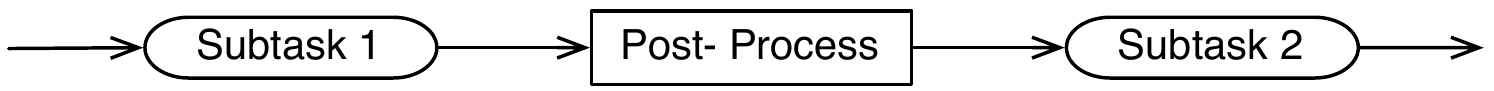}
\vskip-1.5em
\end{figure}


\spara{Problem.}
A complex task generates high latency and low quality output, and needs to be divided into separate parts.

\spara{Solution.}
Two or more crowdsourced processing elements can be composed on a serial, parallel, or hybrid circuit. The output from one CSPE is post-processed and sent to another CSPE (that can also be post-processed automatically). This is related to the now-centenarian paradigm of scientific management/Taylorism/Fordism.
One example can be the decomposition of the task of counting calories in pictures of meals described in~\cite{noronha_2011_platemate}, where a large task (identifying nutrients in a certain food plate) is decomposed to several smaller sub-tasks (drawing boxes to identify the distinct food parts in a plate's image, tag and describe the identified parts, and measure their size in terms of portions). These sub-tasks are handled in a complex workflow within the original crowdsourced task, by different CSPEs, while additional PEs exist to help the process (for example calculating the level of agreement among workers). Similar examples where a large crowdsourcing task is divided into smaller sub-tasks and handled by separate CSPEs include the detect-fix-verify paradigm in~\cite{bernstein2010soylent}, or a secondary grading task to control the quality of a primary task~\cite{sorokin_2008_utility}. 


\spara{Applicability.}
This pattern can be applied when the task can be divided in advance into discrete sub-units. If not, this pattern can be extended by following an approach similar to \cite{kulkami_2012_turkomatic}, where workers are involved not only in solving the sub-tasks, but actually designing the workflow process to be followed, by suggesting decompositions of the original task. In this case, control flows need to connect the different crowdsourced processing elements, indicating changes in the tasks (e.g. decomposition) performed by them.  

\subsection{Human processing optional per item}\label{subsec:guaranteed-throughput}

\spara{Problem.}
A minimum throughput or maximum latency needs to be guaranteed, but crowdsourced stream processing elements can not guarantee that same throughput. Humans may become a bottleneck.

\spara{Solution.}
There should be a path of data flows between the application's input and its output that does not pass through any crowdsourced stream processing element. This means patterns such as {\em supervised learning} or other types of parallel connections need to be part of the design.

\spara{Applicability.}
This pattern can be applied when the minimum number of crowdsourcing calls needed in steady state for the application to function is zero. It can be applied after pattern presented in~\ref{subsec:supervised-learning} when an adequate amount of training data has been gathered. 

\subsection{Human processing mandatory per item}\label{subsec:guaranteed-quality}

\spara{Problem.}
A guaranteed level of quality, which can only be attained through human oversight, needs to be attained. 

\spara{Solution.}
Every path of data flows between the application's input and output needs to pass through a crowdsourced stream processing element. One example is the case of a system to deal with reports of disruptive behavior (e.g. in a massive multi-player online game or similar system), where account suspension or other penalties can only be authorized by humans~\cite{hodson_2013_juries}. Additionally, a minimum level of redundancy can be applied in all such elements if multiple paths exist.

\spara{Applicability.}
This pattern can be applied when low latency does not need to be guaranteed. Enforcing that every data element that will be written in the output needs to pass through a crowdsourced component may introduce latency in the system.


\section{Case Study: Classification of Social Media Messages during a Crisis}\label{sec:evaluation-aidr}


In this section, we study a concrete stream data processing problem, showing that neither pure crowdsourcing nor pure stream processing are satisfactory alternatives. Next, we show how the framework we have described can guide the CSP system design and serve as an analytical tool for its evaluation. Naturally, this real-world example does not use each and every one of the elements we have introduced, but it touches several aspects that are common to a variety of cases.

\subsection{Problem Statement}

In times of crises caused by natural hazards (such as floods and earthquakes), or by human intervention (such as civil unrest or war), people are increasingly turning to social media platforms such as Twitter.\footnote{\url{http://twitter.com/}} In these platforms, users share information related to what they experience, what they need, what they witness, and/or what they consider important to repeat from other sources such as radio or television~\cite{hughes2009twitter}. Social media usage ``rises during disasters as people seek immediate and in-depth information''~\cite{fraustino_2012_start}. 

Several types of user are interested in this information~\cite{hughes_2014_smem}, and their needs are varied. Members of affected communities need to take decisions about how to best protect their lives and property. Government and non-governmental response agencies need as much information as they can get in order to increase the effectiveness of their efforts. Different emergency response agencies are interested in different types of messages during different phases of an emergency~\cite{roy2013tweet4act}. For instance, reports of damage to infrastructures should be directed to some agencies, while reports about shortages of water and/or food should be directed to others.\footnote{The United Nations organizes its agencies into clusters: \url{http://business.un.org/en/documents/6852}.}

In this case study, we use Twitter platform to perform real-time acquisition and classification of users generated content posted online during crisis situations.

\subsection{Crowdsourcing solution} \label{crowdSolution}

Manual classification of messages is not possible given the scale of information that flows on Twitter, even under very generous assumptions.

To estimate the speed with which volunteers can solve these tasks, we use data obtained by Olteanu et al.~\cite{olteanu2014crisislex} during the labeling of six crisis situations using paid crowdsourcing workers.
The simplest task was ``Indicate if tweets are informative for decision makers and emergency responders,'' with several classification options including ``negative consequences,'' ``donations or volunteering,'' ``advice, warnings and/or preparation,'' among others (6 options in total). The crowdsourcing workers are divided into two groups: \emph{trusted} and \emph{untrusted}, depending on whether their judgments agree with that of the authors in a set of test questions (known as golden data). The average time per label for trusted workers was between 7 and 13 seconds, with an average of 9 seconds. Trusted workers were not allowed to label more than 250 items.

This means that one worker can produce 400 labels per hour, assuming s/he can keep that efficiency for such a long period. If we require 3 labels per item in order to have at least some degree of worker redundancy (recommended in a crowdsourcing setting to ensure quality), 
then a crowd of 3,000 volunteers will be needed to label 400,000 items per hour.

In a recent experience with digital volunteers during a large Typhoon in the Philippines in November 2013, the peak number of volunteers was 500, recruited through a number of channels including a large digital humanitarian network.\footnote{\url{http://irevolution.net/2013/11/13/early-results-micromappers-yolanda/}} The peak throughput we can expect from such a crowd is in the order of 67,000 items per hour, or 1,100 tweets per minute.
However, the largest documented peak of tweets per minute during a natural hazard that we are aware of is 16,000 tweets per minute.\footnote{During Hurricane Sandy in 2012: \url{http://www.cbsnews.com/8301-205_162-57542474/social-media-a-news-source-and-tool-during-superstorm-sandy/}} \emph{With pure crowdsourcing, we may be a factor of 10 slower than needed}.


\subsection{Streaming solution}
\label{subsect:streaming}

The automatic approaches to process data streams are indispensable yet challenging to adapt. The infinite length and evolving nature of data streams 
 introduces concept-drift and concept-evolution issues, also known as the \emph{domain adaptation problem} (an automatic machine-learned classifier trained with data from one situation does not perform well in another situation, as also discussed in Section~\ref{sec:intro}). In order to incorporate the underlying changes which have occurred in the streams, incremental updating of the classification model is required by labeling new items, incorporating them as new training data, and learning a fresh model.

In particular for the crisis domain, although crises have some elements in common, they also comprise distinct elements which make domain adaptation difficult, and thus the automatic classification using pre-existing training data is not a satisfactory solution. For example, in~\cite{imran2014coordinating}, authors performed experiments on different crisis datasets, namely the Joplin tornado, the Sandy hurricane, and the Oklahoma tornado, which all struck different areas of the US in 2011, 2012 and 2013 respectively. The performance (AUC)\footnote{AUC: area under the receiver operator characteristic (ROC) curve, a metric used for binary classification. AUC is measured in the [0,1] scale. The AUC score of a perfect classifier is equal to 1.} of different transfer scenarios was 0.52 (train on Joplin, test on Sandy), 0.56 (train on Joplin, test on Oklahoma) and 0.53 (train on Sandy, test on Oklahoma). Similar problems have been identified in another work~\cite{imran_2013_practical}. This findings suggest that crisis-specific training data are needed since they lead to higher accuracy compared to training data from past disasters. 
\emph{With pure automated stream processing, the quality achieved may be significantly low due to the domain adaptation problem.}


\subsection{A Crowdsourced Stream Application: AIDR}


The purpose of AIDR (Artificial Intelligence for Disaster Response),\footnote{\url{http://aidr.qcri.org/}} is to filter and classify in real time messages posted in social media during humanitarian crises, such as natural or man-made disasters~\cite{imran2014aidr}.
Specifically, AIDR collects crisis-related messages from Twitter\footnote{\url{http://twitter.com/}} (``tweets''), asks a crowd to label a sub-set of those messages, and learns an automatic classifier based on those labels. It also improves the classifier as more labels become available.

AIDR users begin by creating a collection process by entering a set of keywords that will be used to filter the Twitter stream. Next, they define the ontologies to be used to classify messages, selecting them from a set of pre-existing ones, or creating them from scratch. AIDR then instantiates an ongoing crowdsourcing task to collect training labels, which are used to train an automatic classifier that runs over the input data. Finally, an output of messages sorted into categories is generated, which can be collected and used to create crisis maps and other types of reports or in real-time decision-making.

\subsubsection{Design overview}\label{aidr-design}

The design of AIDR follows the meta-model described in Section~\ref{sec:framework}.
At a high-level, the application uses the design pattern ``Human processing'', optional per item (Section~\ref{subsec:guaranteed-throughput}), according to which items (Twitter messages) are able to traverse the application independently of the availability of crowdsourcing workers.

\begin{figure}[t]
 \centering
    \includegraphics[width=\columnwidth]{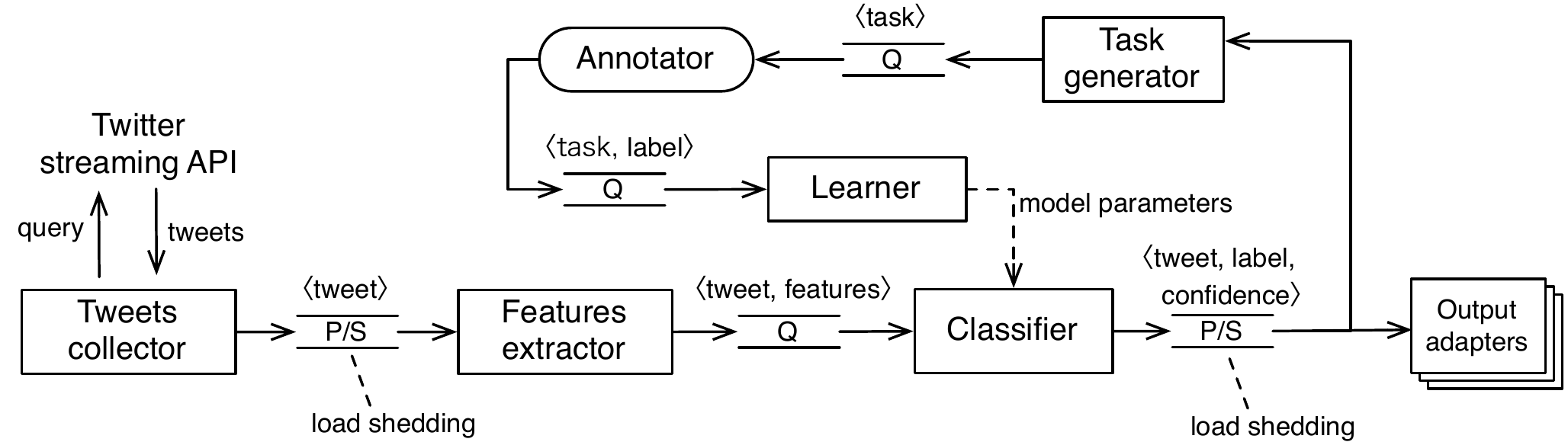}
  \fcaption{AIDR general architecture. {\tt P/S} indicates a publish/subscribe channel, while {\tt Q} indicates a communication queue.}
  \label{fig:aidr-classify}
\end{figure}

The processing elements of the application are composed following the diagram in Figure~\ref{fig:aidr-classify}.
The automatic processing elements include a Twitter collector, feature extractor, task generator, learner and classifier. A crowdsourcing processing element, the annotator, gathers human-provided labels. 
The {\em collector} is an information source automatic processing element that performs edge adaptation~\cite{turaga2010design} to consume tweets using the Twitter streaming API.
The {\em feature extractor} is an information process automatic processing element that prepares the messages by converting them to a set of textual features using standard text operations (extraction of unigrams, bigrams, part of speech classes, etc.).

The classifier, task generator, annotator and learner processing elements {\em interact} following the ``supervised learning'' design pattern (Section~\ref{subsec:supervised-learning}).
The task generator samples the input stream, optionally performing de-duplication (to diversify the elements to label) and active learning (selecting elements for which the classification confidence with the current model is low). 
Both operations are implemented in a stream-aware way, in which the search for near-duplicates and low-confidence elements is done on a bounded-size buffer containing only the latest tweets consumed by the system.
The learner keeps 20\% of the labels it receives as a test set, and uses the remaining 80\% for training, allowing it to report to the user the quality of the current classifier. The learner creates a new classification model (random forest in this case) every 50 training labels and transfers it to the classifier processing element using control signals.

The composition of the processing elements is done through publish/subscribe channels and queues following the hybrid composition model. The channels are able to broadcast and distribute data items, as well as to perform {\em load shedding} - discard elements in order to maintain throughput. The queues are capable of buffering data items, when necessary.

\spara{Implementation.} The design of AIDR follows the {\em Service-Oriented Architecture} software architecture design pattern. All modules depicted in Figure \ref{fig:aidr-classify} provide necessary RESTFul services enabling user interactions for the management purposes such as configurations, providing input, meta-data exchange. Moreover, the communication (in terms of real data) between modules happens using publish/subscribe channels and queues. \mbox{REDIS}\footnote{\url{http://redis.io/}} is used as an underlying communication infrastructure that supports pub/sub channels and queues.

The Java programming language and the Springs 3.0 framework is used for the main application logic, and ExtJS framework~\footnote{\url{http://www.sencha.com/products/extjs/}} for the application user-interfaces. PyBossa~\footnote{\url{http://pybossa.com/}} is used as the crowdsourcing platform. The task generator modules passes newly selected items to be labeled to the PyBossa in order to employ crowdsourcing workers to label items.

\subsubsection{Evaluation}

We apply the evaluation metrics described in Section~\ref{sec:principles}. We remark that there are application-specific evaluation criteria that are outside the scope of our generic framework, e.g., user satisfaction in the case of interactive applications. Here we are interested in the performance (e.g. in high throughput and low latency) and quality (e.g. high classification accuracy of tweets) of using the AIDR CSP system. 

We simulate the operation of AIDR with a fixed input stream under a varying set of streaming conditions. We note that we use real data in our simulation and mimic different streaming conditions to test the system's performance. 
%
For the purposes of this simulation, the application is instrumented to provide detailed information about data passing through it. {\em Mock objects} are also created to simulate data input (from Twitter's API) and data labeling (from a crowd).

We use a dataset of 206,764 tweets containing the hashtag {\tt \#Joplin} and posted during the tornado that struck Joplin, Missouri in 2011~\cite{imran_2013_practical}. We also use 4,000 human-provided labels obtained via CrowdFlower. The specific crowdsourcing task was to indicate if a message is {\em informative} with respect to the disaster and of interest to a broad audience, or if it is either entirely of a personal nature or irrelevant for the disaster (the two latter classes are merged into a single class {\em not informative}).

\begin{figure}
\subfigure[Throughputs: features extractor, classifier, system.\label{fig:throughputs}]{\includegraphics[width=\columnwidth]{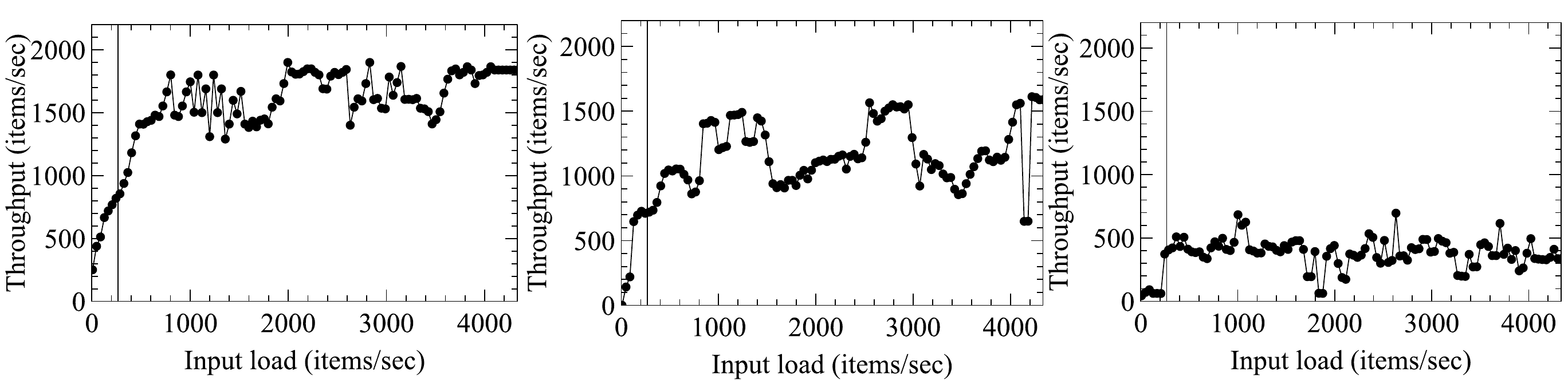}}
\subfigure[Latencies: features extractor, classifier, system.\label{fig:latencies}]{\includegraphics[width=\columnwidth]{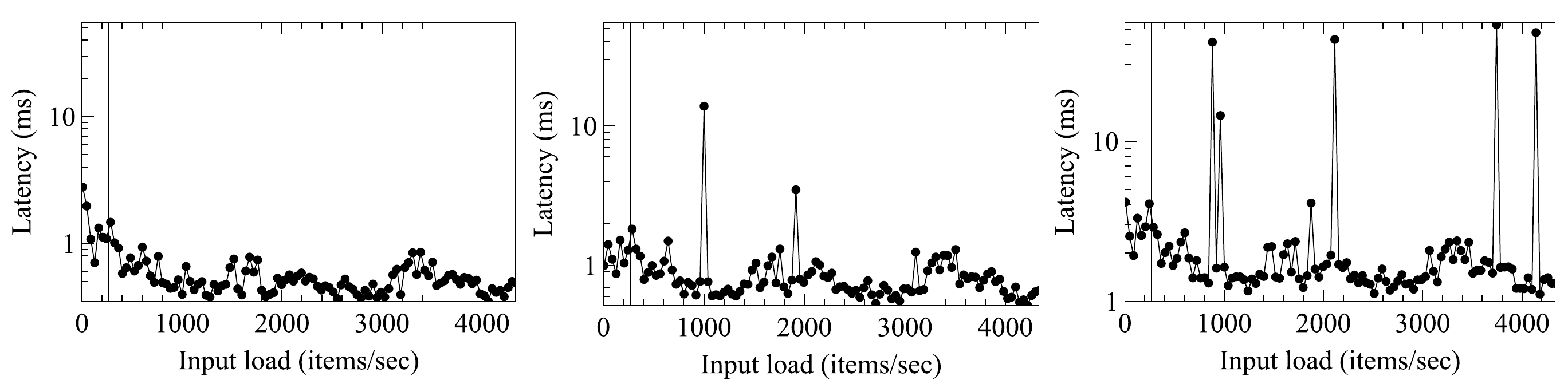}}
\fcaption{Response of AIDR to varying input loads, in terms of throughput and latency. The vertical line corresponds to the highest peak rate documented for a natural disaster to date: 270 items/second during Hurricane Sandy in late 2012.}
\label{fig:eval-stream}
\end{figure}

\spara{Throughput, latency and load adaptability.}
We first measure the attributes that AIDR inherits from being a stream processing application. 
To evaluate these variables, its classifier is trained using all the available labeled data, and then varying input loads are applied.

Figure~\ref{fig:eval-stream} shows the results. In addition to end-to-end throughput and latency, we include a breakdown for each of the two main automatic processing elements of the system: the feature extractor and the classifier.

The system is designed following the principles of {\em keep data moving}, {\em high performance}, {\em task automation}, and {\em load shedding}, outlined in Section~\ref{sec:principles}. The results indicate that the system is able to maintain a high throughput (500 items/second or more) above the observed peak rates in real disasters ($\approx$270 items/second).
Because some of the items are dropped, the latency is kept in the order of tens of milliseconds, even when the input load exceeds the maximum output throughput. For the purposes of this application, this is acceptable given the large amount of redundancy in Twitter messages---about 1/3 are {\em re-tweets} in this dataset, and about half of the remainder are near-duplicates of another message.

\begin{figure}
\centering
\subfigure[Quality vs. number of labels, using passive learning.\label{fig:qua-w-dd}]{\includegraphics[width=0.85\columnwidth]{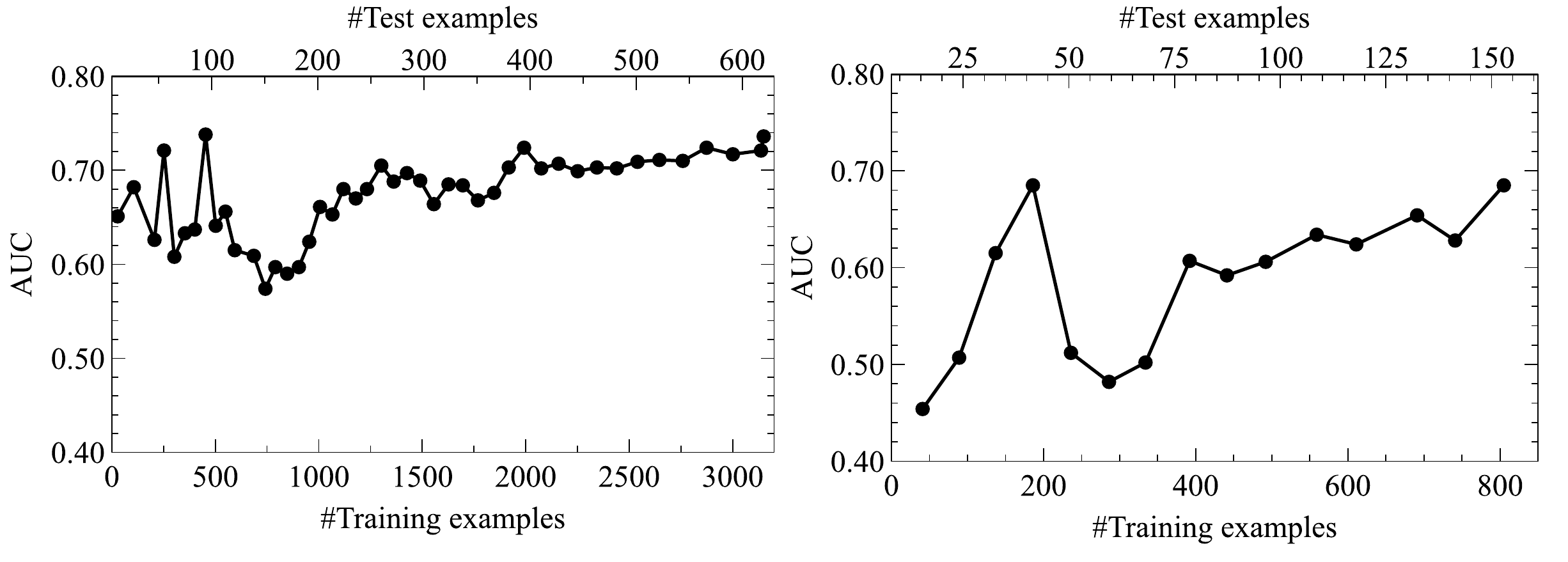}}
\subfigure[Quality vs. number of labels, using active learning.\label{fig:qua-wo-dd}]{\includegraphics[width=0.85\columnwidth]{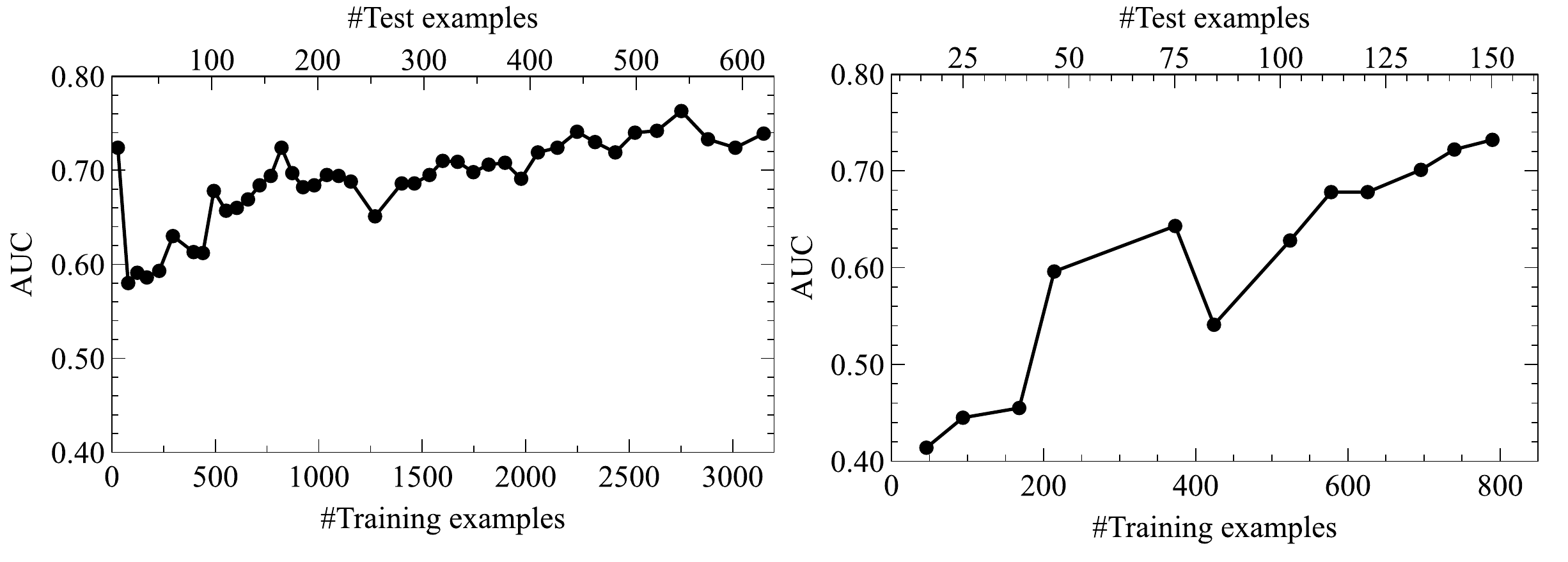}}
\fcaption{Area under the ROC curve (AUC) shows classification quality vs. number of training labels shows cost (i.e. monetary in case of paid-workers, time in case of volunteers). The plots on the left side are obtained without de-duplication, while the plots on the right include it.}
\label{fig:eval-crowd}
\end{figure}


\spara{Quality and cost.}
We next measure the attributes that AIDR inherits from being a crowdsourcing application, specifically, the relationship between quality and cost. Quality is measured using the area under the ROC curve (AUC); higher values are best and 0.5 indicates a random classifier. Cost is directly mapped to the number of human labels used.


Twitter data contains a significant amount of duplicates (i.e. re-tweets). Removing duplicates could decrease the crowdsourcing efforts (i.e. cost) needed to achieve an acceptable level of accuracy. To check this, we are interested in testing two variants of the AIDR system: i) how much cost (i.e. in terms of labeling effort) AIDR requires to achieve an acceptable quality using the passing learning approach and ii) the same using active learning approach, both with and without employing the de-duplication process (i.e. to discard duplicate tweets). In total, we examine four configurations of the task generator: (1) passive learning, equivalent to uniform random sampling in this case, (2) passive learning removing near-duplicate elements, (3) active learning, and (4) active learning removing near-duplicate elements.
Results are shown in Figure~\ref{fig:eval-crowd} where we plot output quality vs. number of labels used by the classifier for training.

The system is designed following the principles of {\em simple tasks} and {\em task prioritization}. The test shows that even under conditions of {\em task frugality} the output quality is acceptable.
After enough training data has been collected ($\approx$1,000 labels), the AUC fluctuations stabilize and the system performs at an AUC of above 0.70. This point is reached with about half the labels if de-duplication is done and it collectively requires 1 minute of 500 workers, if they work simultaneously (as reported in Section \ref{crowdSolution}). Further labels continue to increase the quality of the classifier, with diminishing returns.
Active learning as implemented in this setting (over a bounded-size buffer) does not seem to yield significant improvements.

We remark that in Figure~\ref{fig:eval-crowd} every point represents a different (growing) testing set, a consequence of the online nature of the process.
An offline analysis, where the testing set is fixed to 1/3 of the labeled elements, is in general consistent with the evaluation in the online setting. In the best case and using de-duplication, we obtain a maximum AUC of 0.64 for passive learning (after $\approx$270 labels) and an AUC of 0.66 (after $\approx$200 labels) for active learning.
The offline analysis also shows that more labeled items further improve the results: after using 2,600 labels we reach an AUC of 0.76, which is approximately equal to 3 minutes effort of 500 workers (as reported in Section~\ref{crowdSolution}).


Thus, \emph{compared to a pure stream processing system, the AIDR CSP can be expected to achieve much higher quality} (AUC of 0.76, higher than pure SPs as explained in Section~\ref{subsect:streaming}.)

On the other hand, using a pure crowdsourcing solution, one can achieve AUC 1.00 (by definition, because we are measuring quality against their labels) using a certain level of agreement among workers; however, to achieve such a high accuracy in a pure-stream processing system, we would need 500 workers continuously working for the duration of the event (which may span several hours or days). This is not feasible due to worker burn-out (performance deterioration, lower quality over time) and drop-out (less people over time).
Instead, in a CSP system, to maintain high accuracy, we would only need these 500 workers for 3 minutes and then to re-use them for the same period every  few hours--just to update the system (i.e. the supervised classification model in this case) with new training examples. Thus, \emph{compared to a pure crowdsourcing system, the AIDR CSP achieves a high quality demanding much less effort by the crowd workers, thus using the available worker population more efficiently}.



%
%

\mpara{In summary,} this design enables human intelligence to be applied on a data intensive application. As a stream processing system, the presented CSP is able to keep up with the input loads of even large-scale disasters; and as a crowdsourcing application, it is able to use crowdsourcing work in an effective manner.

\section{Related Work}
\label{sec:related-work}

Stream processing and crowdsourcing are vast research areas, so we focus on connecting our research to previous works covering topics closely related to ours.
This includes engineering principles, frameworks, and taxonomies for the predecessors of CSPs, as well as example systems describing key design choices and best practices.

\subsection{Data stream processing} 

Stream processing research covers many disparate fields (e.g., stock market data analysis \cite{demers2006towards}, fraud detection system \cite{schultz2009distributed}, intrusion detection systems \cite{debar2001aggregation}, disaster prediction systems \cite{broda2009sage}), and has passed through a number of stages.
Real-time stream processing systems perform data mining~\cite{sudhakar2012data}, clustering~\cite{aggarwal2003framework, zhou2008tracking}, classification, time series analysis~\cite{lin2003symbolic}, and other decision support tasks~\cite{zhu2002statstream}.
Furthermore, to support the development of application-specific stream processing systems, there are general-purpose platforms such as S4~\cite{neumeyer2010s4} and STORM~\cite{marz_2013_storm} supporting scalable real-time processing.
These systems are designed for high-speed continuous data ingestion, uninterrupted long-running processing, high-throughput, and low-latency.

We have incorporated key performance indicators from the above works into our evaluation metrics (Section \ref{sec:principles}), while our application design framework has been also inspired to some extent by the design of existing general-purpose stream platforms (Section~\ref{sec:framework}).

\subsection{Crowdsourcing}

Two extensive crowdsourcing surveys can be found in \cite{yuen_2011_crowdsourcing, doan2011crowdsourcing}, which examine a variety of crowdsourcing systems -- often under different names such as collective intelligence, human computation, or social systems.

\subsubsection{Crowdsourcing system taxonomies}\label{subsec:relwork-taxonomies}

First, a number of works categorize crowdsourcing systems from an organizational theory point of view, classifying them according to their \emph{business model functionality}.
\citet{geerts_2009_discovering} performs a model-driven categorization, distinguishing the models of crowdcasting, crowdstorming, crowd production and crowdfunding. 
\citet{yukovic_2009_enterprises} categorizes crowdsourcing systems by business-driven objectives, and further distinguish them in regards to their coordination model (marketplace or competitive-based). 
\citet{lykourentzou_2009_collective} distinguishes different user interaction modalities in crowdsourcing: collaborative, competitive, and hybrid.
\citet{saxton_2013_rules} identify nine basic types which span from social financing to citizen media production models; one of their main conclusions is the need for crowd management which as we describe can be provided by a combination of processing elements, through certain design patterns.

Regarding \emph{type-based classification}, crowdsourcing systems are classified per type of task (simple, complex, creative) in \cite{schenk_2011_practices}, while the different functions of the crowd (crowd rating, creation, processing and solving) are described in \cite{geiger_2011_crowdsourcing}. The first work covers ``what'' crowdsourcing workers do, and the second one covers ``how'' they do it, which is 
in line with the description of the role of humans in CSPs (Section~\ref{sec:taxonomy}). 

There are other works that do not explicitly aim at a taxonomical ordering of crowdsourcing, but contain taxonomical and design elements. \citet{quinn_2011_survey} describe 3 dimensions related to our work: quality control, aggregation, and process order.
The first is a design principle (Section~\ref{subsec:principles-quality}).
The second is an automatic element role (Section~\ref{sec:taxonomy}). 
The third describes interaction models between ``computers, workers and requesters'', and as such it is related to the design patters that we propose (Section~\ref{sec:patterns}). 


\subsubsection{Crowdsourcing frameworks and modeling approaches}

Certain studies propose frameworks and modeling approaches to improve the design of crowdsourcing systems.
\citet{bozzon2013reactive} introduce a reactive crowdsourcing modeling approach focusing on the dynamic control of the crowd, by transforming high-level specifications (e.g. regarding task planning or worker handling) into elementary task-type executions.
\citet{roy2013crowds} propose SmartCrowd, a framework to interactively optimize three crowdsourcing processes: task generation, worker-to-task assignments and task evaluation, by taking into account the uncertainties introduced by the human factor.
A number of works provide goal-driven design principles, which focus on the optimization of crowdsourcing systems for a global-level performance target. In this line, \citet{lykourentzou_2013_guided} present a stepwise modeling approach for the design of corporate crowdsourcing systems, realized in five decision-making and application steps (define goals, characterize jobs, profile workers, identify constraints and design a crowdsourcing optimization algorithm).
Finally, \citet{boutsis_2013_crowdsourcing} focus on the process of task-to-worker allocation, and present a framework comprising four components (task management, dynamic assignment, profiling and scheduling), which aims at guaranteeing efficient crowdsourcing system performance under dynamic conditions of the crowdsourcing environment.

These works basically describe methodologies to optimize specific objectives of crowdsourcing system design. In contrast, our work describes a broad set of objectives and principles in a top-down manner (Section~\ref{sec:principles}), providing a framework against which specific design methodologies can be implemented by composing different elements (Section~\ref{sec:framework}). 

\subsection{Use Cases for CSPs}
%

To the best of our knowledge, no prior work attempts to provide a general framework for engineering CSPs. However, there are works that describe CSP applications (without that specific name). We have already mentioned a number of these works in Section~\ref{sec:taxonomy}.
Concrete examples include web table matching \cite{fan2013hybrid}, entity resolution \cite{whang2013question}, or iterative text recognition \cite{little_2009_turkit}. A common element
in them are high-throughput and low-latency requirements, which we capture in Section~\ref{sec:principles}.
There are also works describing real-time crowd-involving systems (``flash crowds''~\cite{kittur2013future}), such as
the system by~\citet{mashhadi_2011_quality} on quality control for user-contributed data in ubiquitous applications,
the ones proposed by \citet{bernstein2010soylent} and \citet{bigham2010vizwiz},
or the crisis response system described by \citet{rogstadius_2011_realtime} (the latter has been recently adapted to operate with AIDR, the application we describe as a use case in Section~\ref{sec:evaluation-aidr}).

An indicative example is CDAS~\cite{liu2012cdas}, a CSP system for data analytics applied to tasks of sentiment analysis and image tagging. CDAS includes a quality assurance mechanism that takes into account the performance deviations of crowdsourced processing by monitoring historical performance data to estimate each elements' accuracy (instantiation of the quality evaluation metric of Section~\ref{subsec:principles-quality}), as well as an online strategy to reduce waiting time (latency evaluation metric in our framework).  
Finally, systems using a mix of crowdsourced and automatic elements to achieve more efficient treatment of homogeneous data have been described, with two key examples described in
\cite{wang_2013_transitive, franklin2011crowddb}.

Certain elements of our framework borrow notions from the success and limitations showcased in the above systems. As an example, the study of performance in terms of quality, cost and speed, mentioned in the above works, have been expanded into the evaluation metrics (Section~\ref{sec:principles}).
Furthermore, several of the best practices illustrated above have been converted into design patterns (Section~\ref{sec:patterns}).
%


\section{Conclusions}\label{sec:conclusions}

Crowdsourced stream processing is a new computational frontier that combines high-speed processing with human intelligence.
Its progress hinges upon the development of efficient algorithms, as well as on the design of software architectures that implement those algorithms into applications having impact on the real world.
This paper introduces a generic framework that covers system-level properties and behavior, design principles, structural elements, compositions and patterns, to enable better designs for future CSP applications and to serve as a basis for the evaluation and re-engineering of existing ones.

Much remains to be done between the design of specific CSP applications solving concrete problems, and the development of general frameworks and best practices that can serve as a basis for those designs. 
%
This includes extending specialized taxonomies of CSPs, creating new metrics for their evaluation, expanding a catalog of design patterns, among many other tasks that remain open for future work.

\nonumsection{Acknowledgements}
The authors would like to thank Jakob Rogstadius and Ji Kim Lucas for their contributions in the development of AIDR.

\bibliographystyle{nourlabbrvnat}
\bibliography{paper}



\end{document}